\documentclass{ws-ijmpa}

\begin{document}

\markboth{Stephen Godfrey}
{$P$-wave Charm Mesons as a Window to the $D_{sJ}$ States}

\catchline{}{}{}{}{}
%
%%%%%%%%%%%%%%%%%%%%%%%%%%%%%%%%%%%%%%%%%%%%%%%%%%%%%%%%%%%%%%%%%%%%

\title{$P$-WAVE CHARM MESONS AS A WINDOW TO THE $D_{sJ}$ 
STATES\footnote{Supported in part by the Natural Sciences and 
Engineering Research Council of Canada}
}

\author{\footnotesize STEPHEN GODFREY%\footnote{Email: 
%godfrey@physics.carleton.ca}
}

\address{Ottawa-Carleton Institute for Physics, Department of Physics,\\
Carleton University, Ottawa, Canada K1S 5B6 }

\maketitle

%\pub{Received (Day Month Year)}{Revised (Day Month Year)}

\begin{abstract}
In my talk I discussed the properties of the newly discovered 
$D_{sJ}^*(2317)$, $D_{sJ}(2460)$, $X(3872)$, and SELEX $D^*_{sJ}(2632)$
states and suggested 
experimental measurements that can shed light on them.  In 
this writeup I concentrate on an 
important facet of understanding the $D_{sJ}$ states, 
the properties of the closely related $D_0^*$ and $D_1'$ 
states. These states are well described 
as the broad, $j=1/2$ non-strange charmed $P$-wave mesons.

\end{abstract}

\section{Introduction}	

The last sixteen months has seen the discovery of the 
$D_{sJ}^*(2317)$\cite{Aubert:2003fg}, 
$D_{sJ}(2460)$\cite{Besson:2003cp}, $X(3872)$\cite{Choi:2003ue}, 
and $D_{sj}(2632)$\cite{Evdokimov:2004iy} states.  All of these states 
have properties significantly different from what was predicted 
beforehand for conventional $q\bar{q}$ states. This has led to 
considerable theoretical speculation that these states may be 
something new such as multiquark states or meson-molecules.  Another 
point of view is that conventional $q\bar{q}$ explanations cannot yet 
be ruled out and there are diagnostic tests that should be applied to 
understand the nature of these newly discovered states.  In my talk I 
discussed the $q\bar{q}$ possibilities for these new states 
and the quark model predictions that can be used to test them.  Due to 
length restrictions I will restrict this writeup to new results on the 
$D_0^*$, $D_1'$, and $D_{sJ}$ states and refer the interested 
reader to published work on the $X(3872)$\cite{Barnes:2003vb}
and SELEX $D_{sJ}^+(2632)$\cite{Barnes:2004ay} states.

\section{The $D_{sJ}$ States and Their Nonstrange Partners}

The four $L=1$ $P$-wave mesons can be grouped into two doublets 
characterized by the angular momentum of the light quark: $j=3/2$, $1/2$.  
The $j=3/2$ $c\bar{s}$ states were predicted to be relatively narrow 
and are identified with the $D_{s1}(2536)$ and $D_{s2}(2573)$ states
while the $D_{s0}^*$ and $D_{s1}'$ $j=1/2$ states were expected to
have large $S$-wave widths decaying to 
$DK$ and $D^*K$ respectively\cite{Godfrey:1986wj}.  
Quite unexpectedly the Babar\cite{Aubert:2003fg} 
and CLEO\cite{Besson:2003cp} collaborations discovered two
charm-strange mesons in $B$-decay, decaying to $D^+_s\pi^0$ and 
$D_s^{*+}\pi^0$ 
which were below the $DK$ and $D^*K$ threshold respectively.
Virtually all the theoretical effort has concentrated 
on these states \cite{Colangelo:2004vu}.  However, their nonstrange 
partners can also hold important clues to the puzzle but have 
received almost no attention.  

The measured properties of the $L=1$ charmed mesons are summarized in 
Table 1 along with quark model 
predictions\cite{Godfrey:1986wj,Godfrey:1985xj,godfrey2004}. 
The quark model gives a 
$P$-wave cog that is $\sim 40$~MeV 
too high but the splittings are in very good
agreement with the measured masses.  The width predictions are given 
for the pseudoscalar emission model with the flux-tube 
model giving qualitatively similar results\cite{Godfrey:1986wj}.
We note that Belle\cite{Abe:2003zm} 
and FOCUS\cite{Link:2003bd} measure 
$\Gamma(D_2^{*0})=37\pm 4.0$~MeV and $\Gamma(D_1^0)=23.7\pm 4.8$~MeV 
which are slightly larger than the PDG values.  They attribute the 
difference from older results to taking into account interference with 
the broader $D$ states.  Overall the agreement between theory and 
experiment is quite good.

\begin{table}[h]
\tbl{Comparison of Quark Model Predictions$^{7,9,10}$
to Experiment for the $L$=1 Charm Mesons.}
{\begin{tabular}{@{}ccccc@{}} 
\toprule
State & \multicolumn{2}{c}{Mass (MeV)}  & \multicolumn{2}{c}{Width (MeV)} \\
	& Theory$^a$ & Expt & Theory$^{b,7,10}$ & Expt \\ \colrule
$D_2^*$ & 2460 & $2459\pm 2$ $^c$  & 54 & $23\pm 5$ $^c$\\
$D_1$ & 2418   & $2422\pm 1.8$ $^c$& 24 & $18.9^{+4.6}_{-3.5}$ $^c$ \\
$D_1'$ & 2428  & $2438\pm 30$ $^d$ & 250  & $329\pm 84$ $^d$\\
$D_0^*$ & 2357 & $2369\pm 22$ $^e$ & 280 & $274\pm 32$ $^e$\\
\botrule
\end{tabular}}
$^a$ The $P$-wave cog\cite{Godfrey:1986wj,Godfrey:1985xj} 
was adjusted down 42~MeV. \\
$^b$ Using the masses from column 2. \\
$^c$ Particle Data Group\cite{Eidelman:2004wy} \\
$^d$ Average of the Belle\cite{Abe:2003zm} 
and CLEO\cite{Anderson:1999wn} $D_1'^{0}$ measurements\\
$^e$ Average of the Belle\cite{Abe:2003zm} 
$D_0^{*0}$ and FOCUS\cite{Link:2003bd} 
$D_0^{*0}$ and $D_0^{*+}$ measurements. 
\end{table}

%As has been pointed out for the case of the $D_{sJ}$ states, 
Radiative transitions probe the internal structure of 
hadrons\cite{Godfrey:2003kg,Bardeen:2003kt,Colangelo:2003vg}. 
Table 2 gives the quark model predictions for 
E1 radiative transitions between the $1P$ and $1S$ charm 
mesons\cite{godfrey2004}.  Some 
of these transitions should be observable.  The $D_1^0\to 
D^{*0}\gamma$ and $D_1^0\to D^{0}\gamma$ transitions are of particular 
interest since the ratio of these partial widths are a measure of the 
$^3P_1-^1P_1$ mixing angle in the charm meson sector and a good test 
of how well the HQL is satisfied.

\begin{table}[h]
\tbl{Partial widths and branching ratios for 
 E1 transitions between $1P$ and $1S$ charmed mesons.  
The widths are given in keV unless otherwise noted.  
The $M_i$ and the total widths used to calculate the BR's 
are taken from Table 1.  The matrix elements are calculated using the 
wavefunctions of Ref. 9.
}
{\begin{tabular}{@{}l l c c c c c c@{}} 
\toprule
Initial & Final & $M_i$ & $M_f$ &  $k$ & 
	$\langle 1P | r | nS \rangle $ &  Width  & BR  \\
state  & state & (GeV) & (GeV) & (MeV) & (GeV$^{-1}$) & (keV) & \\
\hline 
$D_{2}^{*+}$ & $D^{*+} \gamma $ & 2.459 & 2.010 & 408 & 2.367 & 57 & 0.25\% \\
$D_{2}^{*0}$ & $D^{*0} \gamma $ & 2.459 & 2.007 & 411 & 2.367 & 559 & 2.4\% \\
$D_{1}^+$    & $D^{*+} \gamma$ & 2.422 & 2.010 & 377 & 2.367 & 8.8 & 
				$5\times 10^{-4}$ \\
	   & $D^{+} \gamma$ & 2.422 & 1.869 & 490 & 2.028 & 58 & 0.3\% \\
$D_{1}^0$    & $D^{*0} \gamma$ & 2.422 & 2.007 & 380 & 2.367 & 87 & 0.5\% \\
	   & $D^{0} \gamma$ & 2.422 & 1.865 & 493 & 2.028 & 571 & 3.0\% \\
$D_{1}'^+$    & $D^{*+} \gamma$ & 2.428 & 2.010 & 382 & 2.367 & 37 & 
				$10^{-4}$ \\
	   & $D^{+} \gamma$ & 2.428 & 1.869 & 494 & 2.028 & 15 & 
		$4\times 10^{-5}$ \\
$D_{1}'^0$    & $D^{*0} \gamma$ & 2.428 & 2.007 & 385 & 2.367 & 369 & 0.1\% \\
	   & $D^{0} \gamma$ & 2.428 & 1.865 & 498 & 2.028 & 144 & 
		$4\times 10^{-4}$ \\
$D_{0}^{*+}$ & $ D^{*+} \gamma$ & 2.357 & 2.010 & 321 & 2.345 & 27 & $10^{-4}$ \\
$D_{0}^{*0}$ & $ D^{*0} \gamma$ & 2.357 & 2.007 & 324 & 2.345 & 270 & 0.1\% \\
\botrule
\end{tabular}}
\end{table}

The overall conclusion is that the quark model describes the 
$P$-wave charmed mesons quite well and   
models invoked to describe the $D_{sJ}^*(2317)$ and  
$D_{sJ}(2460)$ states must also explain their non-strange charmed meson 
partners.

Turning to the $D_{sJ}$ states, the narrow $j=3/2$ 
states are identified with the $D_{s1}(2536)$ and $D_{s2}(2573)$
with their observed properties in good 
agreement with quark model predictions\cite{Godfrey:1986wj,Godfrey:1985xj}.  
The $j=1/2$ states were predicted 
to be broad and to decay to $DK$ and $D^*K$ and were not previously 
observed.  But the $D^*_{sJ}(2317)$ is below $DK$ threshold and the 
$D_{sJ}(2460)$ is below $D^*K$ threshold so the only allowed strong 
decay is $D_{sJ}^{(*)}\to D_s^{(*)}\pi^0$ which violates isospin and is 
expected to have a small 
width\cite{Godfrey:2003kg,Bardeen:2003kt,Colangelo:2003vg}. 
As a consequence, the radiative 
transitions are expected to have large BR's and 
are an important diagnostic probe to understand the nature of 
these states\cite{Godfrey:2003kg,Bardeen:2003kt,Colangelo:2003vg}.  
Although there are discrepancies between the 
quark model predictions and existing measurements they can be 
accomodated by the uncertainty in theoretical estimates of 
$\Gamma(D_{sJ}^{(*)}\to D_{s}^{(*)}\pi^0)$ and 
by adjusting the $^3P_1-^1P_1$ mixing angle 
for the $D_{s1}$ states. As in the case of the $D_1$ states, the 
radiative transitions to $D_{s}$ and $D^*_s$ can be used to constrain
the $^3P_1-^1P_1$ ($c\bar{s}$) mixing angle.  

The problem with the newly found $D_{sJ}$ states are the mass 
predictions. Once the masses are fixed the narrow widths follow.  
My view is that the strong coupling to $DK$ (and $D^*K$)
is the key to solving this puzzle.

%\section*{References}

\end{document}